\documentclass[12pt,preprint]{aastex}
\usepackage{graphics,graphicx,rotating}
\usepackage{natbib}
\usepackage{xspace}
\usepackage{amssymb}
\usepackage{amsmath}
\usepackage{epsfig}

\shorttitle{MAGNETIC FIELD EVOLUTION IN BARRED GALAXIES}
\shortauthors{Kulpa-Dybe{\l}, et al.}

\newcommand\cm{\,\rm {cm}}
\newcommand\erg{\,\rm {erg}}
\newcommand\yr{\,\rm {yr}}
\newcommand\Myr{\,\rm {Myr}}
\newcommand\Gyr{\,\rm {Gyr}}
\newcommand\muG{\, \mu{\rm G}}
\newcommand\kms{\,\rm {km}\,\rm{s}^{-1}}

\newcommand\kpc{\,\rm {kpc}}
\newcommand\vis{\,\rm{cm}^{2}\rm{s}^{-1}}

\begin{document}

\title{GLOBAL SIMULATIONS OF THE MAGNETIC FIELD
EVOLUTION IN BARRED GALAXIES UNDER THE
INFLUENCE OF THE COSMIC -RAY -DRIVEN DYNAMO}
\author{K. KULPA-DYBE{\L}\altaffilmark{1}, K. OTMIANOWSKA-MAZUR\altaffilmark{1}, 
             B. KULESZA-\.ZYDZIK\altaffilmark{1},\\  M. HANASZ\altaffilmark{2}, G. KOWAL\altaffilmark{1,3}, 
             D. W\'OLTA\'NSKI\altaffilmark{2} and K. KOWALIK\altaffilmark{2}
}
\altaffiltext{1}{Astronomical Observatory, Jagiellonian University, ul Orla 171, 30-244 Krak\'ow, Poland}

\altaffiltext{2}{Toru\'n Centre for Astronomy, Nicolaus Copernicus University, 87-148 Toru\'n/Piwnice, Poland}

\altaffiltext{3}{N\'{u}cleo de Astrof\'{\i}sica Te\'{o}rica, Universidade Cruzeiro do Sul, Rua Galv\~{a}o Bueno 868, CEP 01506-000, S\~{a}o Paulo, Brazil}
                      
\begin{abstract}
We present three-dimensional global numerical simulations of the cosmic-ray (CR) driven dynamo in barred  galaxies. We study the evolution of the 
interstellar medium of the barred  galaxy in the presence of non-axisymmetric component of the potential, i.e. the bar. 
The magnetohydrodynamical dynamo is driven by CRs, which are
 continuously supplied to the disk by supernova  (SN) remnants. No magnetic field is present at the beginning of simulations but one-tenth of  SN explosions is a 
source of a small-scale randomly oriented dipolar magnetic field. In all models we assume
that 10\% of  $10^{51}\erg$ SN kinetic energy output is converted  into CR energy.

To compare our results directly with the observed
properties of galaxies we construct realistic maps of polarized radio emission.
The main result is that the CR-driven dynamo can amplify weak magnetic
fields up to a few $\muG$ within a few $\Gyr$ in barred galaxies. The obtained $e$-folding time 
is equal to $300\Myr$ and the magnetic field reaches equipartition at time $t\sim4.0\Gyr$.
Initially, completely random magnetic field evolves into large-scale structures. An even (quadrupole-type)
configuration of the magnetic field with respect to the galactic plane can be observed.
Additionally, the modeled magnetic field configuration resembles maps of the polarized intensity
observed in barred galaxies. Polarization vectors are distributed
along the bar and between spiral arms. Moreover, the drift of magnetic arms with respect 
to the spiral pattern in the gas density distribution is observed during
the whole simulation time. 
\end{abstract}
\section{INTRODUCTION}
\label{S:Intro}
Bars are ubiquitous and occur in all types of disk galaxies, from early to late Hubble types.
In near-infrared images about 70\% of all the nearby  disk galaxies are barred \citep{men-07}. 
Bars are astrophysically important not only because they are very common in disk galaxies, 
 but they also can significantly affect the gas distribution \citep{sel-93} as well as the magnetic field configuration. 
The first systematic observations of polarized radio emission from  a sample of 20  barred galaxies show  that their magnetic field topology
 is significantly more complicated than  in the case of grand-design spiral galaxies \citep{bec-02}. The strongest polarized radio emission 
was detected in NGC~1097 and NGC~1365 \citep{bec-05}. The main magnetic field properties observed in the above barred galaxies can be summarized as follows. 
The polarized  emission is the strongest
in the central part of the galaxy, where the bar is present. In this region  
the polarized emission forms ridges coinciding with the dust lanes along the leading edges of the bar. 
The polarization vectors change quickly their pitch 
angles in the bar region whenever they are located upstream the dust lanes and this results in
depolarization valley where the polarized emission almost vanishes. Near the shear shock areas the regions
of vanishing polarized intensity are also observed. In the outer disk  magnetic field vectors form spiral pattern with the maxima of emission along 
 spiral gaseous arms and in interarm regions. 
The average total (regular and turbulent) magnetic field strength for this sample of barred galaxies is $10\pm3\muG$. On the other hand, the average regular magnetic
field, calculated from polarized radio emission, is equal to $2.5\pm0.8\muG$. The strongest total 
magnetic field is detected in the central starforming regions (about $60\muG$ in NGC 1097) and 
in the radio ridges along the galaxies bars ($20\muG--30\muG$ in NGC 1365). In the spiral arms 
of barred galaxies the total magnetic field is about $20\muG$, while regular $4\muG$.
The spiral shape of the magnetic field, large pitch angles, and the observed strengths of
magnetic fields indicate that the galactic dynamo works in those galaxies. 

The original concept of the fast CR driven dynamo was proposed by \cite{par-92}. 
Several researchers have approached this problem numerically, e.g., \cite{han-04,han-09,otm-09,sie-10}.
The CR-driven dynamo  involves the following processes. CRs are continuously 
supplied to the galactic disk due to  supernova  (SN) explosions. The galactic disk stratified
by gravity is unstable due to buoyancy of the magnetic field and CRs. 
Buoyancy effects induce the formation of magnetic loops in the frozen-in,
predominantly horizontal magnetic fields. The rotation of the interstellar gas causes 
that magnetic field loops are twisted by the Coriolis force.  
Next, due to the fast magnetic reconnection small-scale magnetic loops merge to form 
the large-scale radial magnetic field component. The newly created magnetic field component is 
stretched by differential rotation, what results in amplification of the large-scale 
toroidal magnetic field component. The combined action of the above effects is sufficient to trigger the 
exponential growth of the large-scale magnetic field with timescales 
$140--250\Myr$, which are comparable to the galactic rotation period. 

Three-dimensional (3D) MHD numerical simulations in the shearing-box approximation have shown that
the CR-driven dynamo can exponentially amplify weak magnetic fields up to few $\muG$  within few $\Gyr$
in spiral galaxies \citep{han-04,han-09} as well as in irregular galaxies \citep{sie-10}. What is more, some of  the observed 
magnetic fields properties such as  extended halo structures of the edge-on galaxies or the so called
X-shaped structures \citep{soi-05}, can be explained using the CR-driven dynamo \citep{otm-09}. 
The first complete global-scale 3D numerical model of the CR-driven dynamo in an axisymmetric galaxy has been demonstrated recently by \cite{han-09}.
These simulations have given very interesting results and have shown that the CR-driven dynamo is one of the most promising process responsible 
for amplification and maintenance of galactic magnetic fields.
\section{THE INITIAL CONDITIONS AND INPUT PARAMETERS}
\label{S:INIT}
We investigated the evolution of the barred galaxy using the magnetized fluid approximation governed by the isothermal non-ideal
MHD equations \citep{lan-98}. To make the set of above equations complete we include the CR transport. 
Following \cite{schl-85}, the propagation of CR component in the interstellar medium (ISM)  is described by the diffusion-advection equation
\begin{equation}
\frac{\partial e_{cr}}{\partial t} + \mathbf{\nabla} (e_{cr} \mathbf{v}) = \mathbf{\nabla}
(\hat{K} \mathbf{\nabla} e_{cr}) - p_{cr}(\mathbf{\nabla} \cdot \mathbf{v}) + CR_{source},
\label{eq:crdif}
\end{equation}
where $e_{cr}$ is the CR density, $p_{cr} = (\gamma_{cr}-1)e_{cr}$ is the CR pressure,
$\hat{K}$ is the diffusion tensor, $\mathbf{v}$ is the velocity and $CR_{source}$ is the source term for cosmic
ray energy. We assume that CR energy is added to the system by SNe explosions and 
that 10\% of  $10^{51}\erg$ SNe kinetic energy output is converted  into CR energy, while the thermal energy
is  neglected. The adiabatic index $\gamma_{cr}$ for the CR fluid is 
set to be $14/9$. Additionally, CRs are weightless, thus they only contribute to the total pressure (not to the total mass) and are included in the gas motion equation as $\nabla p_{cr}$ \citep{ber-90}. Following \cite{ryu-03}, the anisotropic diffusion of the CR gas is described by diffusion tensor $\hat{K}$ as
\begin{equation}
K_{ij} = K_{\perp} \delta_{ij} + (K_{\parallel} - K_{\perp})n_i n_j,
\end{equation}
where $K_{\parallel}$ and $K_{\perp}$ are parallel and perpendicular (with respect to the local
magnetic field direction) CR diffusion coefficients and $n_i =  B_i/B$ are components 
of the unit vectors tangent to magnetic field lines.

All numerical simulations were  performed with the aid of the Godunov code \citep{kow-09}. 
We numerically investigated the CR-driven dynamo model in 3D in a computational box of the size $30\kpc \times 30\kpc \times 8\kpc$, 
with a spatial resolution of 320$\times$ 320 $\times$ 80 grid zones in the $x$, $y$, and $z$ directions, respectively. The speed of sound is set to $c_s=5\kms$ and  
the initial gas density at the galactic center $\rho_0$ is equal to $1.0$ at~H$\cm^{-3}$.  Following several detailed reviews of the theory of CR diffusion
 \citep[e.g.][]{str-07}, the values of the CR diffusion coefficients assumed in the simulations are $K_{\parallel}=3\times10^{28}\vis = 100 \kpc^2\Gyr^{-1}$ 
and  $K_{\perp} =3\times10^{26}\vis = 1 \kpc^2\Gyr^{-1}$. The resistivity coefficient is the same as in our previous work \citep{kul-09} and it is set to be
$\eta =3\times10^{25}\vis = 0.1 \kpc^2\Gyr^{-1}$, while $\beta = p_{cr}/p_{gas}$ is constant and equal to $1$ in the initial condition. 
At the beginning of the calculation the  magnetic field is not present. 
Following \cite{han-09}, the magnetic field is added  to the galactic disk through randomly distributed SN explosions in the period
 of time $0.1\Gyr--1.1\Gyr$.
During this time interval weak $10^{-5}\muG$ and dipolar magnetic field is supplied in 10\% of SN remnants \citep{han-09}. 
 In the presented simulations the probability of a single SN event is proportional to the local gas density. 
After $t=1.1\Gyr$ the dipolar magnetic field is not longer injected because, due to the dynamo action, its contribution starts to  be insignificant. 
We apply the outflow boundary conditions on external domain boundaries. 

Our model of the barred galaxy consists of four components: the large and massive halo, the central 
bulge, rotating disk of stars, and the bar. They are represented by  different analytical gravitational potentials:
the halo and the bulge components are described by two Plummer spheres,  stellar disk is represented by the isochrone gravitational potential, while
the bar is defined by the prolate spheroid \citep{bin-87}. The bar component is introduced into the galaxy gradually in time, until it reaches
its final mass $M_{bar}$ (from $t=0.1$~Gyr to $t=0.4$~Gyr). In order to conserve the total mass of the modeled galaxy we reduce
the bulge mass, so we have $M_{bar}(t)+M_b(t)=const$ during the calculations. The bar rotates with constant angular speed
$\Omega_{bar}$.  All quantities which characterize the model of barred galaxy are summarized in Table~\ref{tab:bar_params}.   
\begin{table}
\begin{center}
\begin{tabular}{lllll}
\hline
Parameter 	& Name                                  & Value               & Units  \\
\hline
$M_d$       	& Disc mass                   & $4.0 \times 10^{10}$ & $M_{\odot}$ \\
$a_d$ 		& Length scale of the disk    & $0.6$               & $\kpc$ \\
\hline
$M_b$           & Bulge mass                  & $1.5 \times 10^{10}$ & $M_{\odot}$ \\
$a_b$       	& Length scale of the bulge   & $5.0$               & $\kpc$ \\
\hline
$M_h$       	& Halo mass                   & $1.2 \times 10^{11}$ & $M_{\odot}$ \\
$a_h$       	& Length scale of the halo    & $15.0$              & $\kpc$ \\
\hline
$M_{bar}$       & Bar mass                    & $1.5 \times 10^{10}$ & $M_{\odot}$ \\
$a_{bar}$       & Length of bar major axis    & $6.0$               & $\kpc$ \\
$b_{bar}$       & Length of bar minor axis    & $3.0$               & $\kpc$ \\
$c_{bar}$       & Length of bar vertical axis & $2.5$               & $\kpc$ \\
$\Omega_{bar}$  & Bar angular velocity        & $30.0$              & $\Gyr^{-1}$ \\
\hline
$R_GB$          & Galaxy radius	              & $13.5$ 		    & $\kpc$ \\
CR              & Corotation radius           &	$6.0$	            & $\kpc$ \\
IILR            & Inner Inner Lindblad Resonance  &$0.4$            & $\kpc$ \\
OILR		& Outer Inner  Lindblad Resonance  &$3.0$	    & $\kpc$ \\
OLR		& Outer Lindblad Resonance    &	$8.5$	            & $\kpc$ \\
\hline
\end{tabular}
\end{center}
\caption{Adopted Parameters for the Barred Galaxy Model}
\label{tab:bar_params}
\end{table}
\smallskip
\section{Magnetic field evolution}
\label{S:RESULT}
%
%
\begin{figure*}
\begin{center}
\includegraphics[width=0.5\columnwidth]{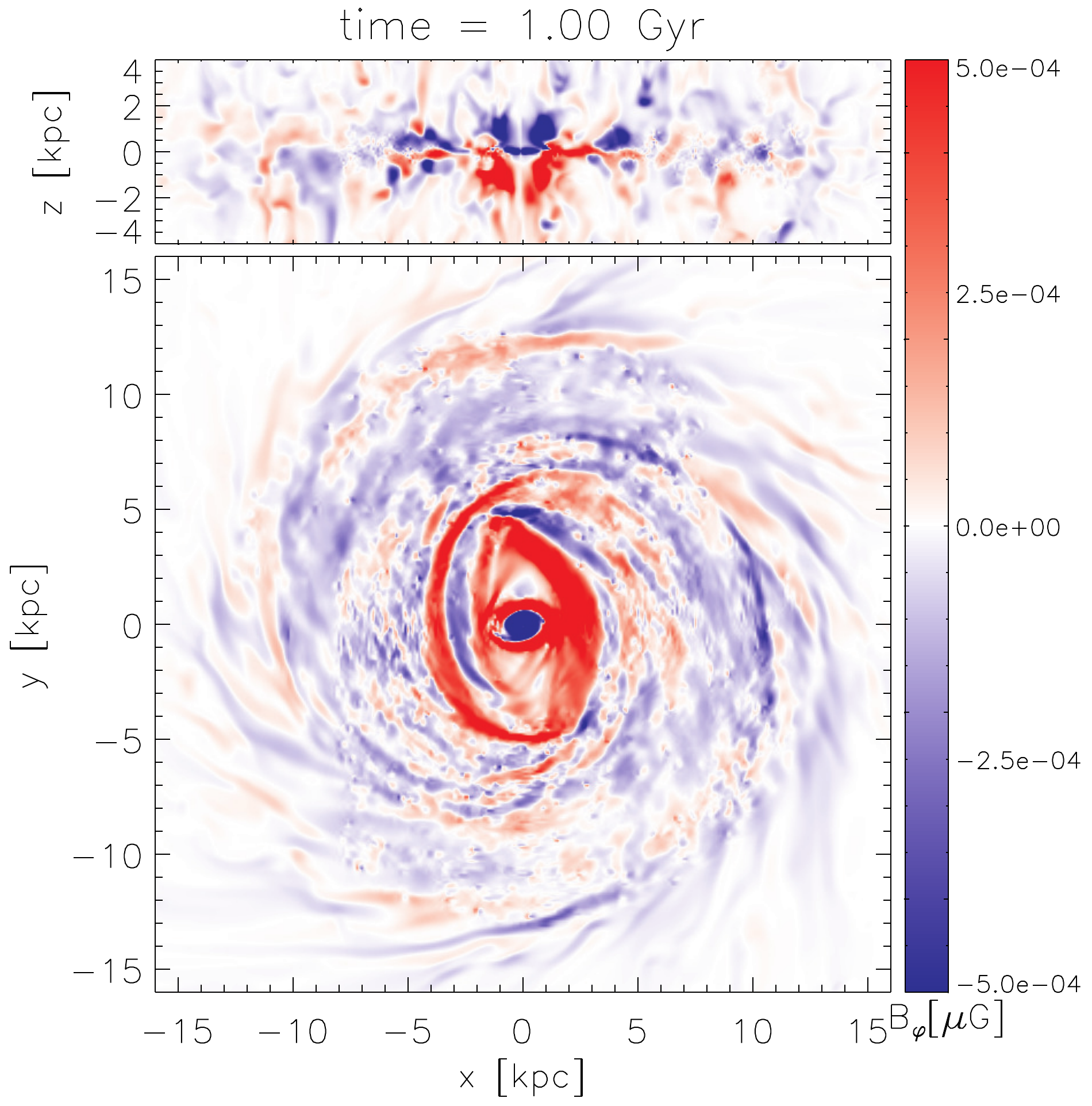}\includegraphics[width=0.5\columnwidth]{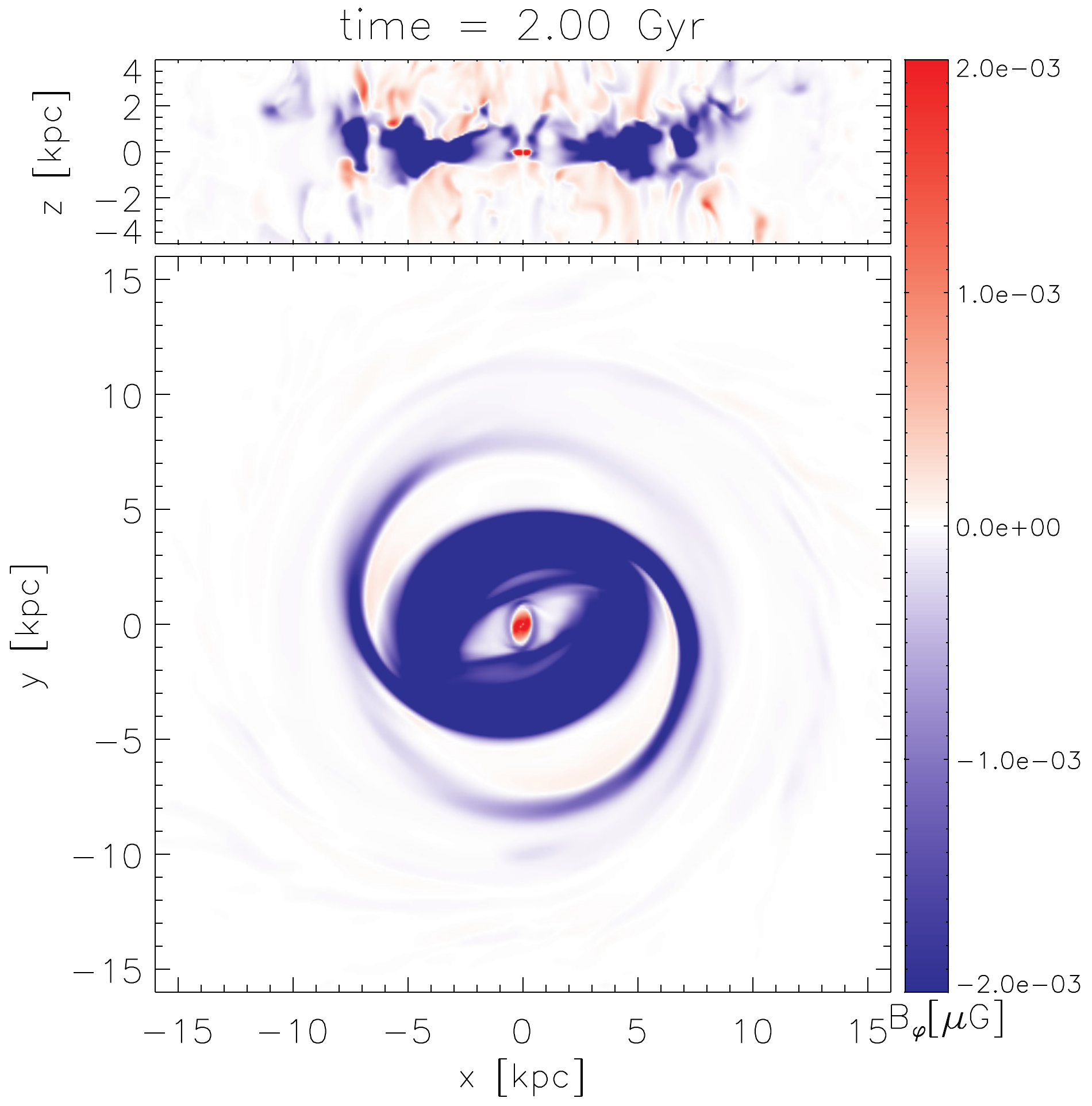}\\
\includegraphics[width=0.5\columnwidth]{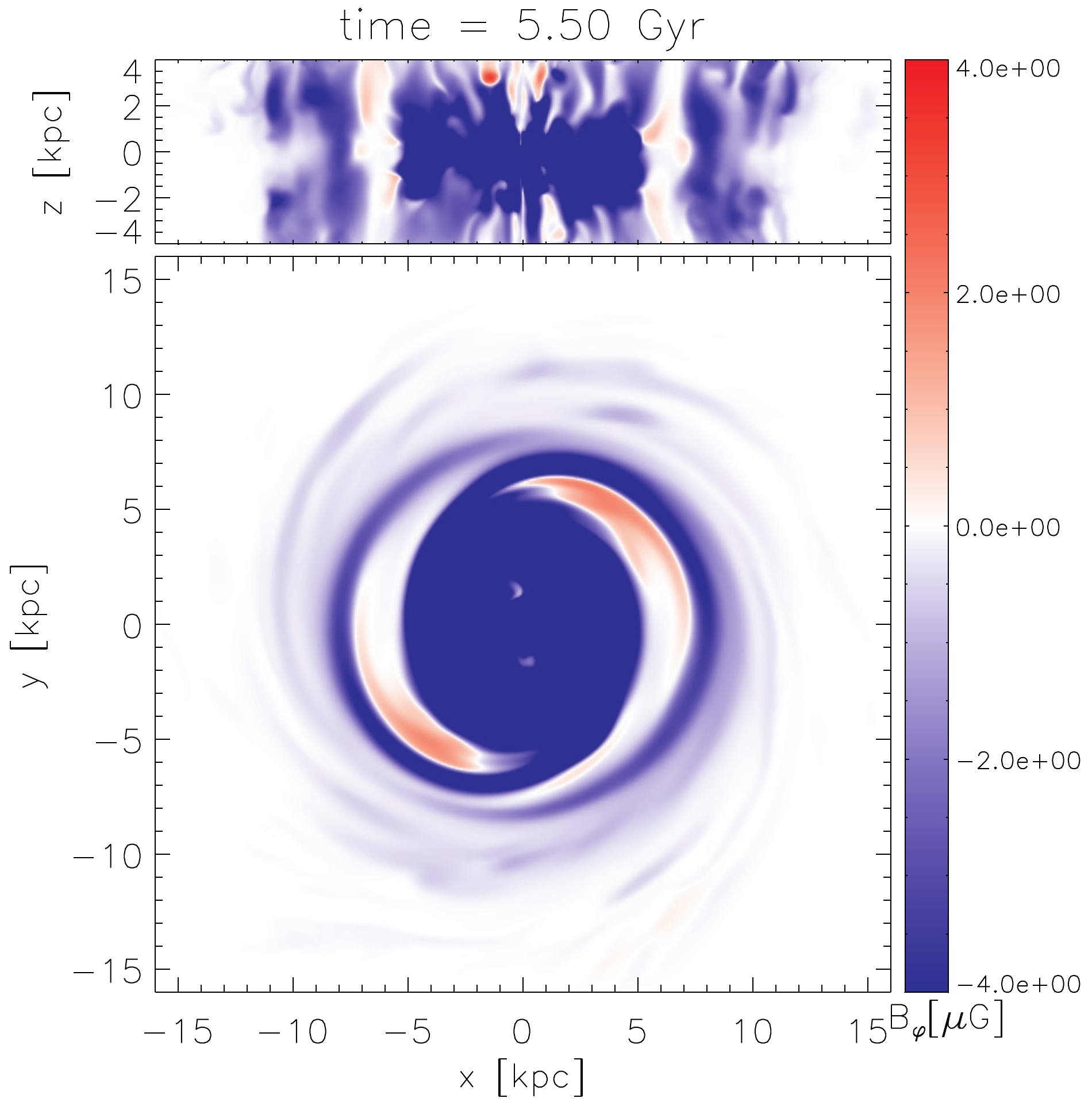}
\caption{
 Distribution of the toroidal magnetic field in vertical and horizontal slices through the disk center
 for selected time steps. Red color represents the regions with the 
positive toroidal magnetic field, blue with negative, while unmagnetized regions of the volume are white. To enhance 
weaker structures of the magnetic field in the outer galactic disk (e.g., magnetic arms), the color scale in magnetic field maps is saturated.
\label{fig:bphi_maps}
}
\end{center}
\end{figure*} 
%
%
\begin{figure*}
\begin{center}
\includegraphics[width=0.5\columnwidth]{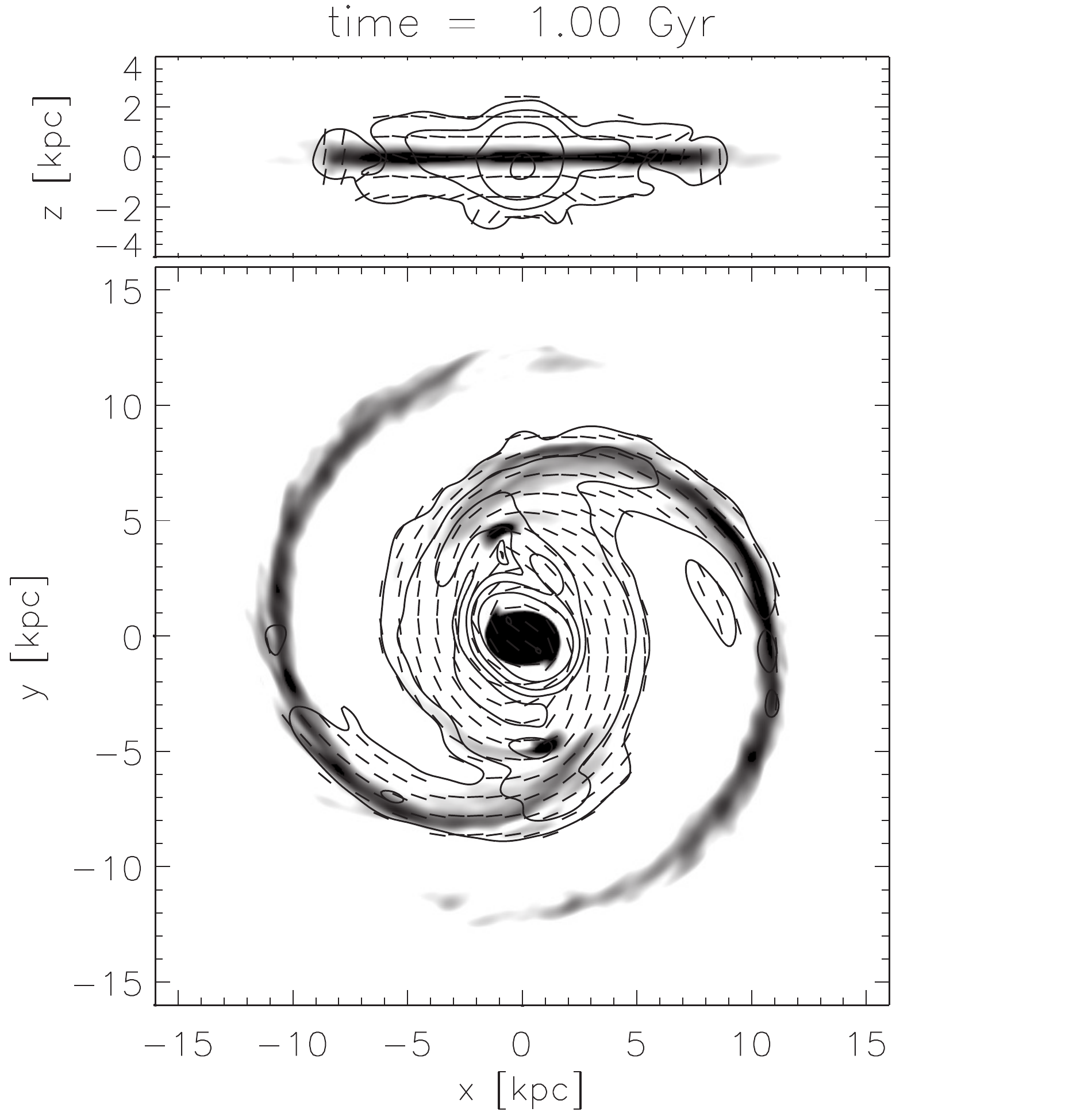}\includegraphics[width=0.5\columnwidth]{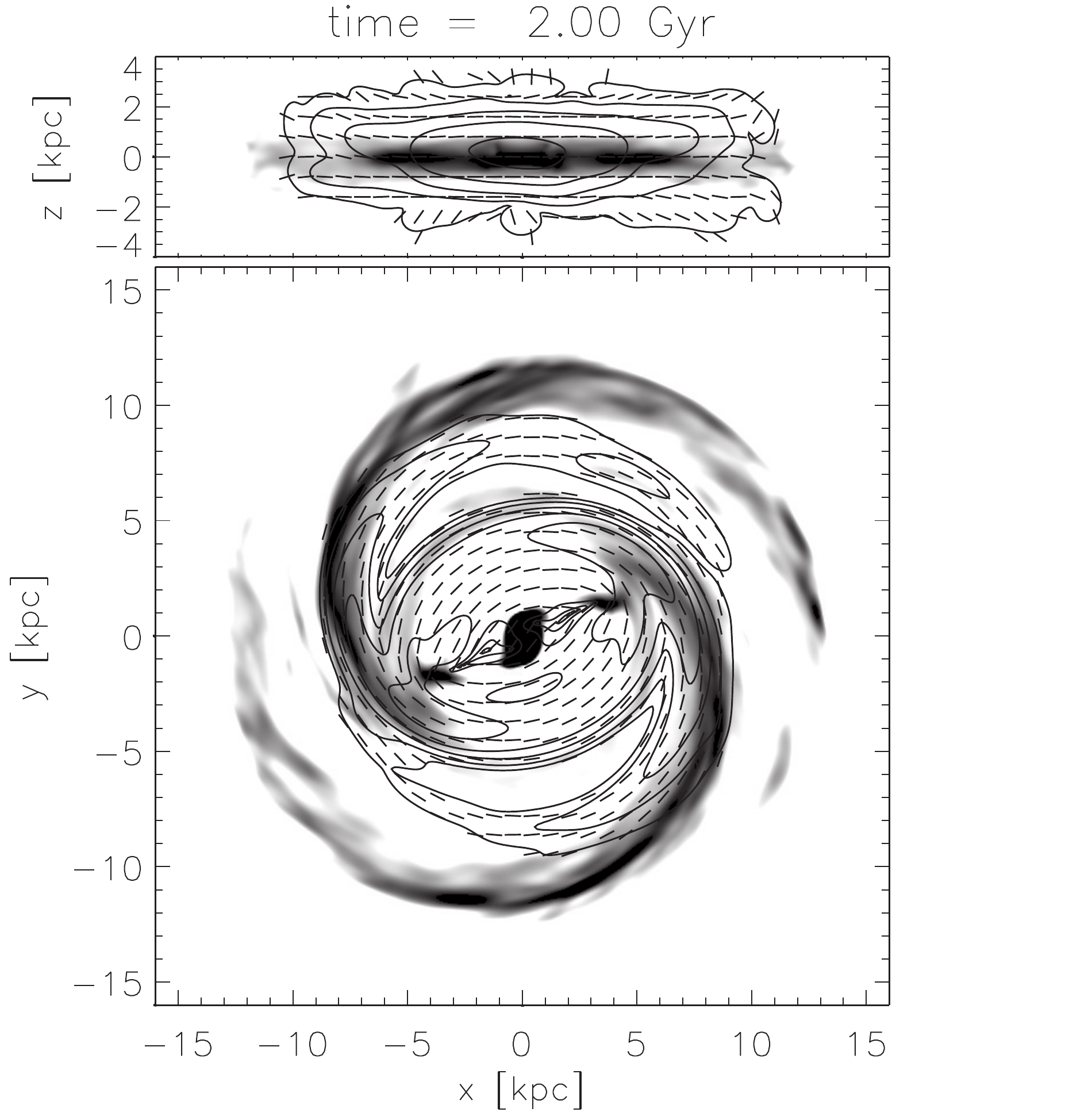}\\
\includegraphics[width=0.5\columnwidth]{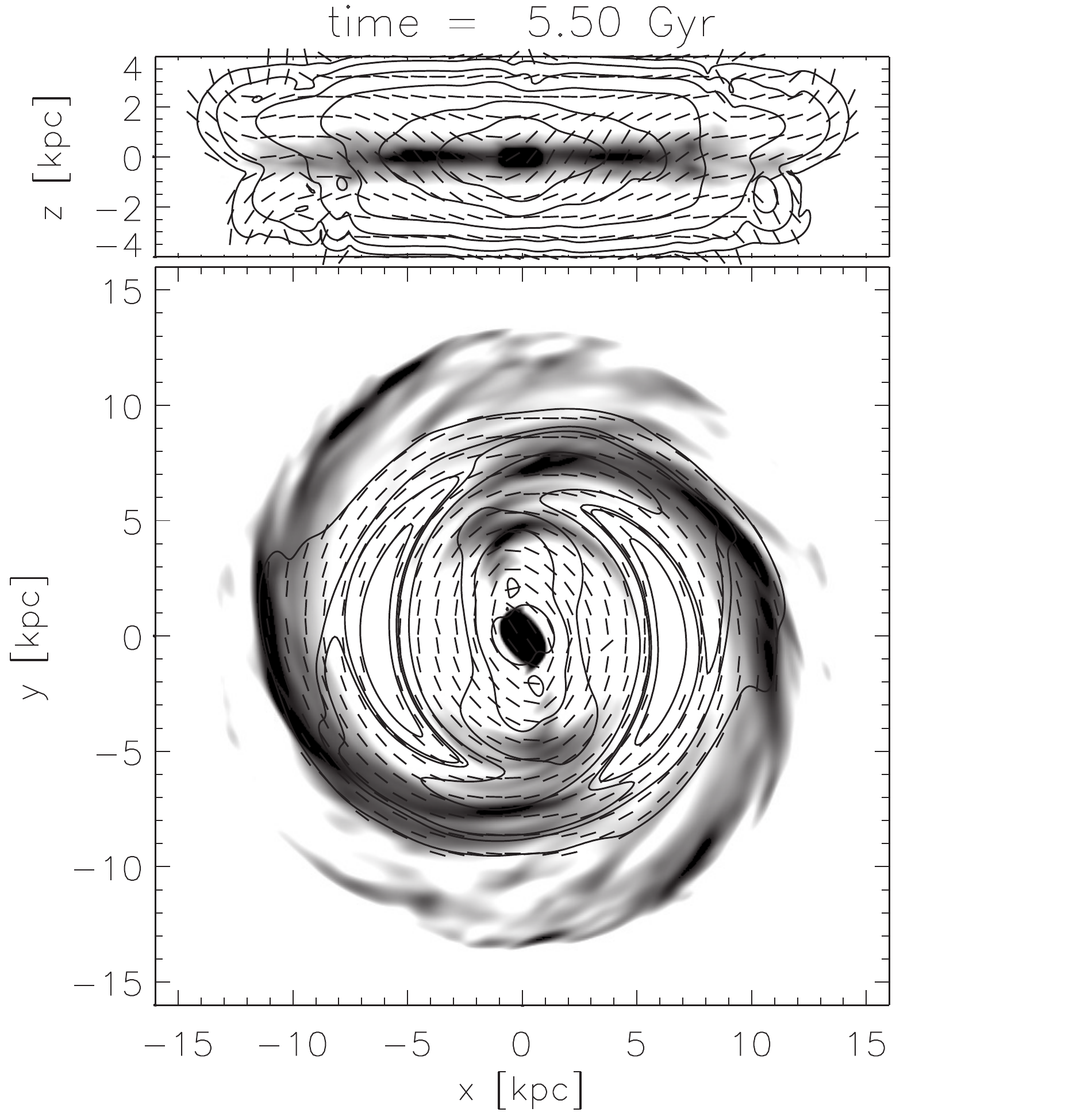}
\caption{ Face-on and edge-on polarization maps  for selected time steps. Polarized intensity (contours) and polarization angles (dashes) are superimposed onto column
density plots.
\label{fig:pol_maps}
}
\end{center}
\end{figure*}
In Figure~\ref{fig:bphi_maps} the toroidal magnetic field component in horizontal and vertical slices  is plotted. Red color represents the regions with the 
 positive toroidal magnetic field, blue with negative, while unmagnetized regions of the volume are white.
The random magnetic field component dominates initially (up to about $t\sim1.0\Gyr$), 
as it  originates from randomly oriented  magnetic dipoles. The ordered magnetic field is visible in 
the inner part of the galaxy,
where it follows the gas distribution, namely the bar and dust lanes. At later epoch ($t=2.0\Gyr$) the toroidal magnetic field component forms well-defined magnetic
 arms which can be observed till the end of the simulation. 

The positive toroidal magnetic field component
gradually reaches higher values, both in the bar and magnetic arms, and occupies larger volume of space in the galactic disk and halo.
 On the other hand, the gas motions in the bar region  generate shocks which enhance  the magnetic field in this area
in addition to the ongoing dynamo process. 
Indeed, the regular magnetic  field in magnetic arms at time $t=5.5\Gyr$ is equal to $7.3\muG$,
while in the bar region the regular magnetic field reaches  $50.1\muG$.
 
Reversals of the magnetic field can be observed at time  $t=2.0\Gyr$ (Figure~\ref{fig:bphi_maps}) in the very inner part of the bar, where the negative toroidal magnetic field  
component  is present.
However, these reversals  disappear almost completely in subsequent epochs. Additionally, due to the influence of the nonaxisymmetric gravitational potential, magnetic reversals 
are apparent  between the bar and magnetic arms at time $t=5.5\Gyr$ in Figure~\ref{fig:bphi_maps}.  

At the beginning of the calculation the randomly distributed toroidal magnetic field dominates what is apparent in vertical slices. 
Next, at time  $t=1.0\Gyr$ (Figure~\ref{fig:bphi_maps}),
  the odd (dipole-type) configuration of the magnetic field with respect to the galactic plane can be observed. However, this configuration is transient and after 
 $t=2.0\Gyr$ in Figure~\ref{fig:bphi_maps} an even (quadruple-like) symmetry of the magnetic field dominates.
The quadrupole-like symmetry was also obtained by \cite{han-09} who studied the CR driven dynamo in an axisymmetric galaxies.
 Moreover, small
 reversals appear during the whole simulation time.  

In Figure~\ref{fig:pol_maps} we present the magnetic field evolution in synthetic polarization maps  for the same epochs as in Figure~\ref{fig:bphi_maps}. 
The polarization maps show the distribution of the polarization angle and polarized intensity superimposed onto the column density.
 All face-on and edge-on polarization maps have been smoothed down to the resolution $40''$.
 The dark shades  represent regions of the highest density.
The magnetic field maxima correspond to the gas density enhancement, 
where the SN explosions are located, what can easily be seen at  $t=1.0\Gyr$ (Figure~\ref{fig:pol_maps}). At this time step, the magnetic field is present in the gaseous arms 
as well as in the central part of the  galaxy. 
In the shock regions in the bar the rapid change of the magnetic field direction is apparent.
Moreover, no regular magnetic field is observed in the interarm regions, however magnetic arms start to detach from  gaseous spirals.  
At time  $t=2.0\Gyr$ the magnetic spiral is clearly visible between the bar and gaseous arms. 
The drift of magnetic arms is persistent and
takes place during the whole simulation time. Its shape changes slightly 
as the simulation proceeds. For instance, at  $t=2.0\Gyr$ the  magnetic spiral is well 
defined in the interam region, while at  $t=5.5\Gyr$ 
it is less visible because it connects with magnetic structures
apparent in the bar region. 
The drift of magnetic arms into the interarm area was described in a number of papers e.g.,  \citep{kul-09,kul-10}, where the authors concluded that this behavior 
is caused by difference in the angular velocity of magnetic arms and the gaseous spirals. Namely, the magnetic arms do not corotate with gaseous spiral structure, but they follow
the general gas motion in the disk, which has a slightly lower angular velocity.   However, in \cite{kul-09,kul-10} no dynamo action was included
and the drift of magnetic arms into the interam region is not observed during the whole simulation time but only in the short period of calculation.

In the edge-on maps in Figure~\ref{fig:pol_maps}, the polarized synchrotron emission extends far outside the galactic plane.
Near the disk plane the magnetic field 
is mainly parallel do the disk, while in the halo vertical magnetic field components can also be seen. The strong vertical field in the halo is probably transported
 by mass outflow from the galactic disk.  The averaged outflow rate is equal to $1.1M_{\odot}\yr^{-1}$.
The most extended structures are apparent at time $t=5.5\Gyr$. This configuration of the magnetic field vectors bears some resemblance to the extended magnetic
 halo structures of the edge-on galaxies \citep{kra-09}. 
%
%
\begin{figure*}
\begin{center}
\includegraphics[width=0.5\columnwidth]{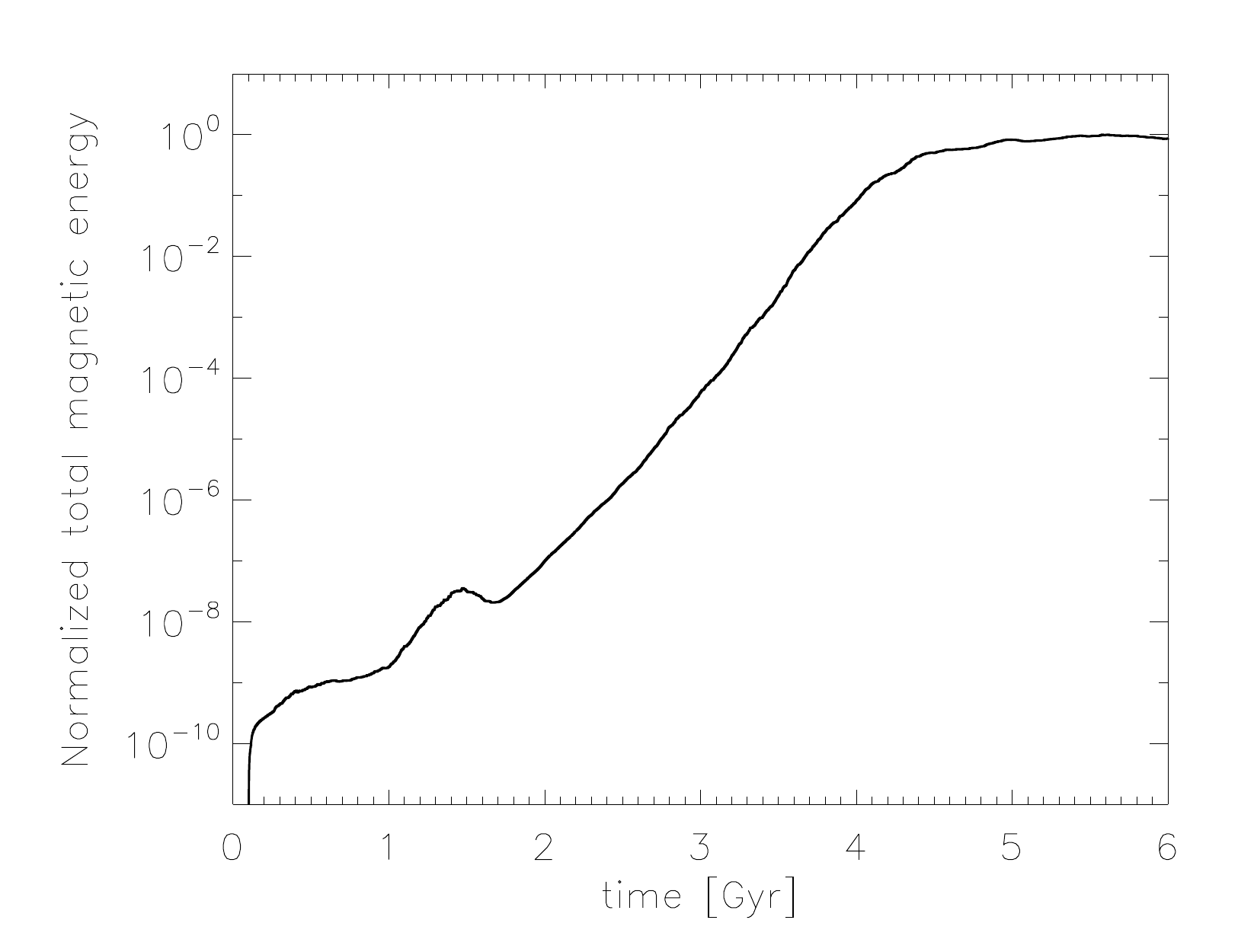}\includegraphics[width=0.5\columnwidth]{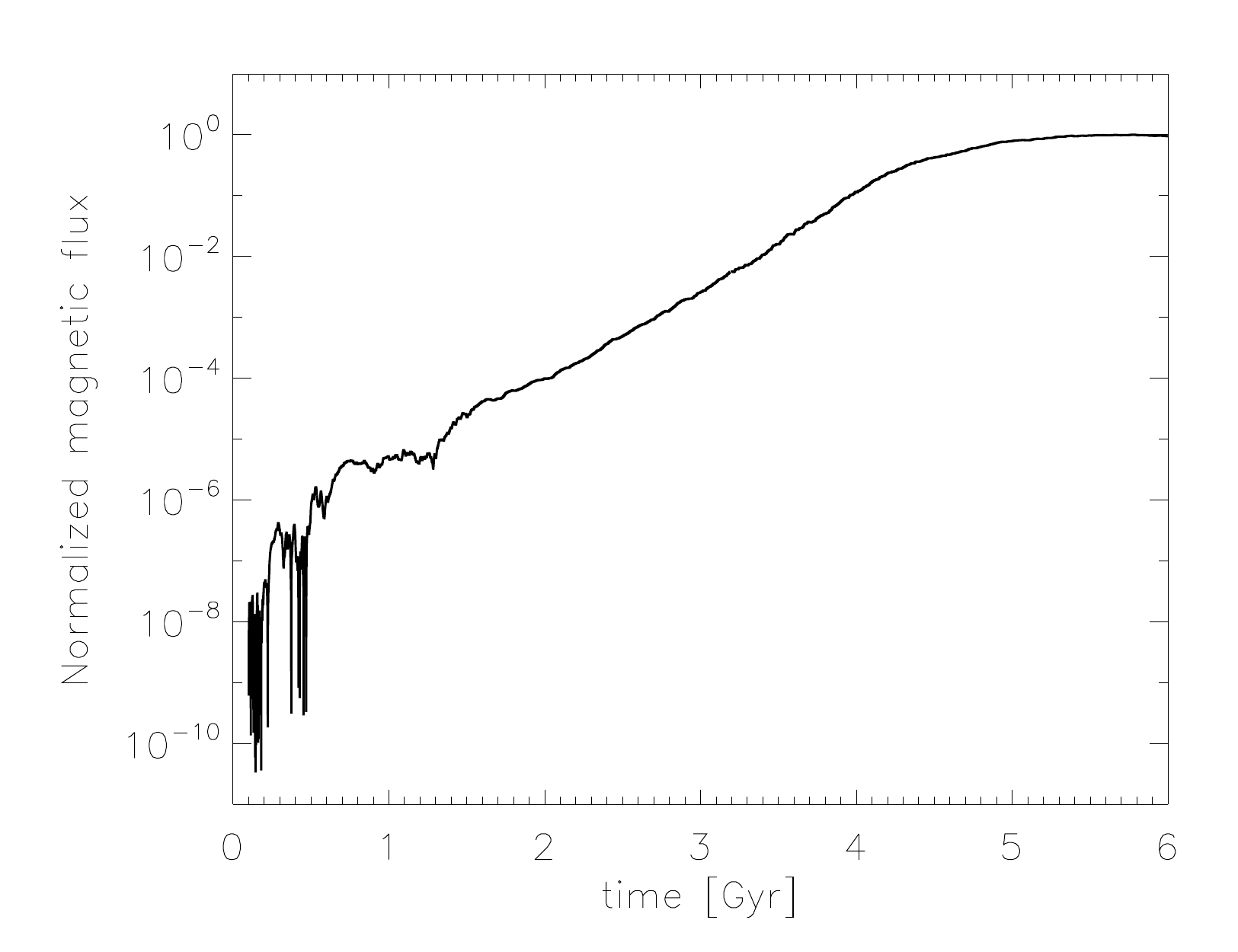}
\caption{
Evolution of the total magnetic energy $E_B$ (left panel) and the mean azimuthal flux $B_{\phi}$  (right panel) for the modelled 
barred galaxy. Both quantities are normalized with respect to the equipartition value.}
\label{fig:emag}
\end{center}
\end{figure*}

The CR-driven dynamo action in barred galaxies causes that the total magnetic field (left panel) and  azimuthal  flux (right panel) plotted in  
Figure~\ref{fig:emag} grow approximately exponentially. The growth of the total magnetic field  lasts until the saturation level is reached at time $t=4.5\Gyr$.
The growth of the regular magnetic field
 is  identified with the amplification of the azimuthal magnetic flux.
The regular magnetic field grows up on an average timescale ($e$-folding time) equal to $300\Myr$. 
After time $t=1.32\Gyr$, the growth of the magnetic flux is clearly exponential and lasts until the equipartition is reached.
\section{Conclusion}
\label{S:Conclusion}
This work demonstrates for the first time the action of the  CR driven dynamo in barred galaxies.  
Many observational  futures of the magnetic field in  barred galaxies have been reproduced.
\begin{itemize}
\item The polarized radio emission is  strongest in the inner part of the bar and in radio ridges that approximately
follow the dust lanes indicated by the enhancement of the gas density.
\item In the outer part of the disk magnetic vectors  form a spiral pattern with  maxima of polarized intensity along spiral gaseous arms and 
in interarm regions.  The drift of magnetic arms into the interarm area is observed during the whole simulation time.
\item The synthetic edge-on radio maps of polarized emission resemble magnetic structures observed in edge-on barred galaxies.
\item The obtained average strength of the magnetic field and the maximum value of the total magnetic field in magnetic arms as well as in the bar  are  consistent with observational values. 
\item The large-scale magnetic field has a quadrupole-like symmetry  with respect to the galactic plane.
\end{itemize}
The CR-driven dynamo is a powerful mechanism for amplifying magnetic fields in barred galaxies. 
During the whole simulation time we see the contribution driven by CRs driving  turbulence, apparent through the buoyant motions.
We incorporate the sub-grid physics through the resistivity of ISM. This amount of resistivity $\eta =3\times10^{25}\vis$
is exactly what is needed to ensure that magnetic energy in small-scale magnetic fields and in large-scale 
magnetic fields are equal in the CR-driven dynamo model \citep{han-09b}.
The corresponding diffusive timescale is the magnetic field-ordering timescale.
The timescale $t_{grow} \sim\ln(B/B_0) l / v$ tells us how much the random magnetic field on scales ($l$ and $v$) 
would grow in the absence of resistivity. But since, due to the resistivity, the conversion of small-scale 
magnetic fields to large scales is efficient, we observe the growth of the large-scale magnetic fields to 
the equipartition values (from the initial $B_0$), exactly in the period of 3--4 Gyr.
\acknowledgements
This work was supported by the Polish Ministry of Science and
Higher Education through grants: 92/N-ASTROSIM/2008/0 and 3033/B/H03/2008/35.
The  computations presented here have been performed on the GALERA supercomputer in 
TASK Academic Computer Centre in Gda\'nsk.
\bibliographystyle{apj}

\end{document}